# Relativistic Paradoxes and Lack of Relativity in Closed Spaces

Moses Fayngold,
Department of Physics, New Jersey Institute of Technology, Newark, NJ 07102

Some known relativistic paradoxes are reconsidered for closed spaces, using a simple geometric model. For two twins in a closed space, a real paradox seems to emerge when the traveling twin is moving uniformly along a geodesic and returns to the starting point without turning back. Accordingly, the reference frames (RF) of both twins seem to be equivalent, which makes the twin paradox irresolvable: each twin can claim to be at rest and therefore to have aged more than the partner upon their reunion. In reality, the paradox has the resolution in this case as well. Apart from distinction between the two RF with respect to actual forces in play, they can be distinguished by clock synchronization. A closed space singles out a truly stationary RF with single-valued global time; in all other frames, time is not a single-valued parameter. This implies that even uniform motion along a spatial geodesic in a compact space is not truly inertial, and there is an effective force on an object in such motion. Therefore, the traveling twin will age less upon circumnavigation than the stationary one, just as in flat space-time. Ironically, Relativity in this case emerges free of paradoxes at the price of bringing back the pre-Galilean concept of absolute rest.

An example showing the absence of paradoxes is also considered for a more realistic case of a time-evolving closed space.

> *Although Einstein's static universe is no longer acceptable now that we know the universe is expanding, the closed universe geometry remains a possibility.*
> Alan Guth

> *…Matter tells spacetime how to curve;*
> *spacetime tells matter how to move.*
> Misner, Thorne, Wheeler

> *Maybe relativity does not like certain topologies…*
> Anwar Shiekh

## 1. Introduction

A closed (or compact) space, just as an open space, is a continuous set without any boundaries or edges, but with all its dimensions being closed loops. A space with at least one open dimension can be named "semi-closed" or "half-open" (e.g., a conical surface).

Certain closed spaces are modeled sometimes by identifying images at the opposite edges of a flat screen, for instance, in a video with a personage leaving one edge to appear on the opposite one. *If all 4 edges are connected by such identification, the video can be considered as taking place on the surface of a torus* [1]. We will not consider such models, to avoid possible ambiguities. Identifying screen's opposite edges in our mind is different from identifying them physically. The former is a mathematical exercise, whereas the latter *demands* physical bending of the screen. If a character leaving one edge of the screen reappears at the opposite edge, his/her motion is not continuous. Physically, one cannot return to the opposite edge without turning back or going into embracing space to make a loop. In this work, we assume a (semi)closed space to be a real physical space and consider some implications.

For the sake of brevity, we denote an open Euclidean space as A, while a (semi)-closed space as B. For their comparison, we assume both – A and B – to have the same number $D$ of dimensions, $D(A) = D(B)$. But we also consider a hypothetical Euclidean space $\tilde{A}$ embedding B, with $D(\tilde{A}) = D(B) + 1$. Space $\tilde{A}$ must be flat in order to expose the topology of B and to show that geodesic in B may be not a geodesic in $\tilde{A}$. An example illustrating this point is a circle $L \in B$, which is a single geodesic of a $1D$ space B. Adding to it a new dimension orthogonal to the plane of $L$ produces a cylinder which is just another B-space ($B \to \tilde{B}$), only semi-closed. It does embed the initial B, but not in a way needed for our goals, since $L$ remains geodesic in it, and its curvature remains non-visualized for its residents. On the other hand, adding radial dimension to B makes it a Euclidean plane which is an eligible $\tilde{A}$-space. $L$ is *not* a geodesic in $\tilde{A}$ – it loses this status even though *it has not been distorted*. Keeping B fixed when invoking $\tilde{A}$ is another crucial requirement in this work.

Altogether, the presented work deals with a version of a problem considered in [2], but in a broader context and using a different approach, applying a simplified model of B-space with $D = 2$ (e.g., a cylindrical or a spherical surface). The results will be sufficient to illustrate the absence of paradoxes also in B with $D > 2$. In a way, such an approach is similar to the one used in [3], but with focus on physical effects.

Physics of B-spaces has many unusual features. For instance, diverging field lines of a charge $Q$ at a point O converge at the antipodal point $O'$. The geometry of the field lines at $O'$ seems identical to that at O, but the net field of $Q$ there is zero even though the distance $OO'$ may be small. The same conclusion follows for a point mass. This effect will be considered in more details in Sec. 10.2.

Some thought experiments with motion along a geodesic will be discussed from the viewpoint of A or $\tilde{A}$. But we will also consider a few thought experiments within B alone, showing that all known relativistic paradoxes have resolution in B as they do in A or $\tilde{A}$.

Almost all examples below deal with static B. It is only weakly relevant to our real time-evolving space. But static B can still be considered as a crude model of some possible states of the universe. For instance, a universe with the initial Big Bang and the average mass density above certain critical value is closed and undergoes expansion followed by contraction, according to some earlier models [4, 5]. The static B can approximate such world during transition between the two stages.

## 2. Statement of the problem: paradoxes in A and B

We start with comparing conditions in A and in B for three most-known paradoxes. And the oscillating universe paradox will be briefly outlined in the end of this section.

### 2.1 *The twin paradox*

*A) The twin paradox in* A

Consider two residents of A – the twins Alan and Alice. Alan is at rest in some inertial frame A1 while Alice makes a round trip. Upon Alice's return she must have aged less than Alan. However, in Alice's view from her frame A2, it is Alan who makes the round trip, and thus he must have aged less. The paradox results from assuming the equivalence of both frames. But Alice must use force to turn around, so A2 is not inertial. There appears a gravitational field in it, whereas A1 remains field-free [4, 5]. So the two frames are not equivalent. This point clarified, both agree that Alice has aged less than Alan.

*B) The twin paradox in* B

Now turn to B-space [2, 6] with two twins – Ben and Bella. Ben is resting in an inertial frame B1, while Bella travels around. In this case Bella can choose to follow a geodesic, e.g., the equator on a 2*D* sphere (Fig. 1). The conventional wisdom is that uniform motion along a geodesic is inertial. Bella returns by virtue of the topology without having to reverse direction; physically - without turning on the engines of her ship. So both frames – B1 and B2 – appear now totally equivalent, and the twin paradox becomes a real paradox. When the twins meet again, Bella is younger than Ben, *and* Ben is younger than Bella.

### 2.2 *The three clock paradox*

*A) The three clock paradox in A*

This is a complicated version of the twin paradox, involving three twins – Alan, Alice, and Ann in space A, and all three in different, generally *non-inertial*, frames. Alan remains in place in a rotating frame A1, say, somewhere on the Earth's equator. Two others depart simultaneously with equal speeds along the equator, but in the opposite directions: Alice – westward (frame A2), and Ann – eastward (frame A3). Despite the apparent symmetry in the traveler's flight conditions, they do not return to Alan simultaneously: Alice will return earlier than Ann. But the readings of their own clocks will be identical. In other words, the traveling clocks return at different coordinate times but read equal proper times. When Alice returns,

Alan will have aged less than her. But when Ann returns, he will have aged more than Ann. Analysis of this experiment is most simple in a non-rotating RF ($A_0$) associated with the center of the Earth. Equal proper times of the traveling twins can be understood qualitatively by noting that Ann moves relative to $A_0$ faster than Alice and must accordingly age slower. But on the other hand, it will take her longer to catch up again with Alan. These mutually opposing factors lead to equal proper times for Alice and Ann. The theoretical predictions have been confirmed already in experiments with non-relativistic velocities requiring very high accuracy [7-9].

B) *The three clock paradox in B*

A similar situation seems to lead to a real paradox in B. Here the twins Ben, Bella, and Bob can move with different velocities along some geodesic. Then each could claim that he/she was at rest, while the other two made a circumnavigation around the Universe, with the result that the claimed stationary twin must have aged more than the travelers upon their return. And each seems to be right since all 3 engines would in this case remain idle all the time.

## 2.3 *The Ehrenfest paradox*

A) *The Ehrenfest paradox in A*

The Ehrenfest paradox deals with two congruent disks of radius $R$ – one stationary (A1), with the rim $A_R(1) = 2\pi R$, and the other – rotating (A2). The rim of A2 (as well as all its inner circles) must be Lorentz-contracted, but since it remains congruent with A1, its *proper* length $A_R(2)$ must exceed $2\pi R$. But on the other hand, in the RF co-rotating with A2 it is $A_R(1)$ that must exceed $2\pi R$ in order to be congruent with $A_R(2)$.

The paradox is resolved when we realize that rotating frame A2 is non-inertial, and thus is not equivalent to A1 (as in case 2.1, it is permeated by a gravitational field not associated with any mass). Also, at any moment in A2, the corresponding time is not a single-valued characteristic [5, 9-14]. While the whole space-time as observed from A2 remains Lorentzian, the geometry of its spatial subset (plane of the disk) is non-Euclidean, and specifically, the proper length of A2 is greater than $2\pi R$:

$$A_R(2) = \gamma(\Omega R) A_R(1) = \gamma(\Omega R) 2\pi R \qquad (1)$$

Here $\gamma(v)$ is the Lorentz-factor, $v = \Omega R$ is the peripheral linear velocity of A2, and $\Omega$ is its angular velocity. Physically, the material constituting A2 is in a state of complex deformation [5] whose origin and mechanism are described in [9, 14]. An important aspect of this phenomenon is that even permanent congruence of two objects does not necessarily imply their equal proper sizes [15]. These factors resolve the paradox: all observers will agree that $A_R(2)$ is given by (1) when measured in A2, but is Lorentz-contracted down to $2\pi R$ in frame A1. And the rim $A_R(1)$ with *proper length* $2\pi R$ is Lorentz-contracted down to $2\pi R / \gamma(\Omega R)$ in A2 while remaining congruent with $A_R(2)$ if we take a proper account of the time lag [9].

B) *The Ehrenfest paradox in B*

However, the paradox appears to resurrect in a compact space, if the rim is a geodesic like the equator on a sphere in Fig.1. Then it seems natural to assume that a uniform sliding along geodesic is an inertial motion equivalent to a state of rest. Denoting the rims of the two equatorial geodesics uniformly sliding along each other as $B_R(1)$ and $B_R(2)$, respectively, different observers will obtain mutually exclusive results. Suppose Ben is on geodesic B1, while Bella slides down it on the congruent geodesic B2. Then Ben insists that the proper length of B2, in order to be congruent with B1, must exceed $B_R(1)$ by the Lorentz factor, $B_R(2) = \gamma(\Omega R) B_R(1)$. But Bella, using the same argument, gets $B_R(1) = \gamma(\Omega R) B_R(2)$.

## 2.4 *The Oscillating Universe paradox*

Consider now a more realistic case of the universe evolving with time as described by the Friedman equations [16]. If the average mass density is above critical, it forms a physical B-space first expanding and then contracting. Such case involves B alone, since A, due to the absence of any scaling parameter, cannot be oscillating.

Suppose that Ben and Bella, now residing separately in two different galaxies of B, synchronize their time via optical signaling. Suppose that the cosmic background radiation turns out to be isotropic in both galaxies. This means that both are stationary in B and thereby with respect to each other. At the same time, they recede from each other due to cosmological expansion, and then get again close together during contraction stage before the Big Crunch. And each twin can claim that he/she was remaining in place all the time while the other made a round trip. Due to time dilation, the moving twin must have aged less than the stationary one. But since each can claim the "resting" status with equal right, we seem to have a real paradox.

## 3. Resolution of paradoxes: general considerations

In all considered cases an implicit assumption has been made that the equivalence of inertial frames in relative translational motion in A holds without restrictions for uniform motions along a geodesic in B. But since this assumption leads to contradictions, it must be wrong. The question arises – what specifically makes it wrong? And if not entirely, then to what degree the relativity of motion holds in a closed space?

Presented below is the analysis of these questions, and possible answers.
One purely formal answer follows from the direct comparison between a closed and a flat space. We can always imagine B imbedded into $\tilde{A}$. Then it is immediately seen that a geodesic in B is not geodesic in $\tilde{A}$ (Fig. 1). Accordingly, the difference between states of rest and motion along such curve becomes obvious. If B is stationary in $\tilde{A}$, but its clone $B'$ is uniformly sliding along B, then B and $B'$, even though permanently congruent, are not equivalent, since "sliding down" here is rotational rather than translational motion. The physical difference between such motions resolves the first three paradoxes.

The resolution of the forth paradox is a little more subtle, but also simple. For each twin, the motion of the other can be observed via the Doppler shift – first red, then blue. Generally, the relativistic Doppler-effect consists of two contributions – one from the change of distance between the light source and detector, and the other from time dilation due to their actual relative motion. It is only the second contribution that leads to the twin paradox. But precisely this contribution is absent in the given case since both twins do not physically move relative to B and to each other, and all their observed motion is only due to expansion and contraction of space itself. This can be visualized, e.g., in a model with 2 ants sitting still on the surface of an expanding balloon. Their changing separation is caused only by the expansion, rather than by their own motion. The absence of the latter means the absence of the corresponding time dilation. This resolves the paradox.

Thus, invoking an embracing space $\tilde{A}$ makes the analysis trivial. But such approach, albeit the simplest, is not unanimously accepted. There are a few conventional arguments against it.

1) As is well known, the geometry of B can be learnt from the inner measurements without peeping out into $\tilde{A}$. Moreover, B may be the single reality of the world (just as 3-D macro-space of our Universe). Then any references to $\tilde{A}$ is merely a speculation.

2) One must be careful in conclusions about B based on its viewing from $\tilde{A}$. For instance, a Klein's bottle intersects itself in $3D$ space but can be not self-intersecting in $4D$ space [17].

3) If B is semi-compact, it can have the zero Riemann curvature, and in this respect is similar to A. And yet it can show paradoxes when we consider motion down a geodesic with non-zero component along a closed dimension.

All these arguments are correct. But they do not prove the assertion that relativistic paradoxes are irresolvable in B.

Objection 1) means that a B-space can be autonomous. But this does not make reference to $\tilde{A}$ illegal. Even if $\tilde{A}$ does not exist physically, we can still use it as a mathematical construct helping to prove or clarify certain properties of B. The above-mentioned model of $2D$ balloon expanding in $3D$ Euclidean space is routinely used in Cosmology to illustrate some features of expansion of the real Universe. And, as we will see below, some physical effects in B *imply* the effective, if not actual, existence of $\tilde{A}$.

Objection 2) uses an assumption that properties of B can be changed by deforming it along the new dimensions offered by $\tilde{A}$. An example simpler than a Klein's bottle is an "8"-shaped curve. It intersects itself in a $2D$ space, but can be made non-self-intersecting in a $3D$ space by moving one of its crossing sides into the 3-d dimension. This will convert "8" into a twisted loop. But we can as well leave it non-deformed; in which case it remains self-intersecting. As emphasized in Sec. 1, when invoking $\tilde{A}$ we do just this – *keep the B fixed*. Then viewing B from $\tilde{A}$ cannot change its physics and geometry.

Objection 3) cannot preclude comparing motions along two different geodesics in B – one with component along the closed dimension, and the other straight. Such comparison reveals that the two motions are not equivalent (a specific example will be considered in Sec. 4).

Actually, the difference between states of rest and uniform motion along a closed geodesic can be seen without invoking $\tilde{A}$, e.g., by considering time synchronization. Suppose we managed to uniquely synchronize all local clocks along the equator in Fig. 1. Then the time in B is a well-defined single-valued characteristic. Consider now another set of equidistant clocks, sliding along the equator. They form a system B′ in the same space, but moving relative to it. Their synchronization leads to the result that at one moment of B-time, the B′-clocks moving ahead of some picked clock $B'_j$ will read local times earlier than that of $B'_j$, and clocks behind $B'_j$ will read later times (Fig. 2). It is immediately seen that B′ does not have a single-valued global time. So there is a fundamental difference between B and B′: the latter is characterized by a specific time lag $\Delta\tau'$ which is absent in the former. In contrast with the length-contraction, which is a local effect, the time lag is generally a global characteristic. But it can be measured without looking out into $\tilde{A}$ [5, 9-14], and all observers agree that $\Delta\tau = 0$ in B and $\Delta\tau' \neq 0$ in B′. The two systems are not equivalent.

Apart from time synchronization, the physical difference between the stationary B and uniformly moving B′, as well as criterion selecting B, can be seen in a simple thought experiment. Suppose Ben sends simultaneously two light pulses in the opposite directions along a closed geodesic (taken as the equator) and they return simultaneously (we leave aside the question how long it may take). Then Ben knows that he is at rest in B.

Let Bella do the same from her East-bound spaceship (frame B′) moving relative to Ben. Even though her motion is uniform and along geodesic, the two light pulses from her ship will not return to her simultaneously: the West-bound pulse will return earlier than the East-bound pulse. The time difference between them is a signature of the non-zero time lag (actually, one determines the other). The only way to get rid of it is to jump to frame B. Bella can do it by boosting her ship in the direction from which the latest of the two signals returns and then sending a new pair of signals to check the outcome. If the time lag still exists, she can slightly boost her ship again and repeat her test. Reiterating this by small increments, she will eventually reach the state at which both signals return to her simultaneously. She can reduce the process to one step by first calculating the necessary boost, and then performing it. The disappearance of the lag will tell her that she is now at rest in B.

Calculating the necessary boost may be easier in $\tilde{A}$, but all the described actions are performed in B which is the only physical space for the B-twins. A unique feature of the whole experiment is its double nature: it is local since it involves a single spaceship; but it is also global by virtue of using circumnavigating photons. The only possible interpretation of this experiment is that space B determines its intrinsic *frame of absolute rest*, and the global system of B-clocks with single-valued time occupies this frame. Any other global B′-system moving with respect to B without distortions is in a state of rotation which differs from rest by emergence of the time lag and specific gravitational field.

Of course, a non-evolving B-space is only of academic interest. In real space, the resolution may be different. It may be impossible to complete the described experiments in the evolving closed Universe. In this case not even photons might be able to finish circumnavigation before the cosmological collapse [6]. This is similar to a local effect with a black hole evaporating a falling object before the fall [18]. In this way Nature can avoid possible contradictions by preventing effects leading to them. The collapse precluding any completed circumnavigations would be another resolution of the twin paradox – physical conditions leading to it would be prevented from happening. In any case, the closed space comes out contradiction-free.

Below we consider some paradoxes and their resolution in more detail.

### 4. Semi-compact spaces

We start with a semi-compact B-space mentioned in statement 3) of Sec. 2. The simplest example is a cylinder. Its surface has the zero Riemann curvature, but non-zero curvature of the closed dimension. Consider two geodesics in this space: one – generatrix OO′, and another – helix OBO′(Fig. 3). All *local* experiments in the vicinity of their intersection O show their equivalence – they are both as good as two intersecting straight lines in a plane. But a *global* experiment around the whole pitch shows essential difference between them. If we simultaneously send 2 light pulses from O along the respective geodesics, they will not arrive at O′ simultaneously. Paths OO′ and OBO′ have different proper lengths:

$$\text{OO}' \equiv \Lambda_A, \quad \text{OBO}' \equiv \Lambda_B = \sqrt{\Lambda_A^2 + (2\pi R)^2} > \Lambda_A, \tag{2}$$

where *R* is the radius of the closed dimension. In this respect $OO'$, by virtue of being shorter, is "more geodesic" than $OBO'$ [1]. One would definitely choose $OO'$ for communications between O and O'.

Consider now a thought experiment with Alan (visiting B from A) and Ben. Let their ships pass O simultaneously – with Alan along $OO'$ and Ben along $OBO'$. Their engines are off, but their respective speeds $v_A$ and $v_B$ are such that they arrive at O' also simultaneously, that is

$$\frac{\Lambda_B}{v_B} = \frac{\Lambda_A}{v_A} = T, \qquad (3)$$

where *T* is their common travel time (coordinate time) measured by a stationary system of clocks. But when the partners meet at $O'$, their individual clocks will read the different proper times

$$\tau_A = T/\gamma(v_A); \qquad \tau_B = T/\gamma(v_B) \qquad (4)$$

In view of (2), (3) we have $\tau_B < \tau_A$ – Ben will have aged less than Alan. Ben cannot dispute this by claiming that his RF was equivalent to Alan's by virtue of being inertial. The distance between Alan and Ben was first increasing from zero to $\pi R$ and then decreasing back to zero, as in the traditional twin paradox. And since Alan's motion is indisputably inertial, such behavior of distance indicates Ben's frame as not inertial. The two frames, even if moving uniformly along their respective geodesics, are not equivalent.

### 5. A compact 2D space

Consider now a totally compact 2D space with constant non-zero curvature. This takes us back to Fig. 1. All geodesics here are the circles of radius *R*. Take one of them as the equator. It is intersection $C_{AB}$ of B and the equatorial plane A, which makes it the common element of A and B. Here we assume that $C_{AB}$ when treated as a geometrical line, has the zero *linear* mass density regardless of whether it is considered as part of A, $\tilde{A}$, or B.

For a flat resident of A, line $C_{AB}$ is not geodesic, and the uniform motion along it is definitely not inertial – the corresponding moving object is under a centripetal force

$$F_A = \mu v^2/R = \mu \Omega^2 R, \qquad (5)$$

where $\mu = \mu_0 \gamma(\Omega R)$ is its relativistic mass. Elimination of A cannot affect the physics of motion along $C_{AB}$. This suggests that B must generate an effective force $\mathbf{F}_B$ of its own, which is orthogonal to B and acts on any *moving* object. The latter property makes $\mathbf{F}_B$ similar (but not identical!) to the Lorentz force. By its action, it is analogous to a normal force on a smooth pebble sliding without friction inside a static balloon along some geodesic. Since $\mathbf{F}_B$ is orthogonal to B, it cannot be directly observed in any local experiment. But, as we have seen in Sec-s 2, 3, it manifests itself indirectly in global experiments within B. We came to a point where physics, instead of discouraging us from peeping out into A or $\tilde{A}$, is asking us to look, at least mentally, beyond B. And it does so through geometry which tells us that B is curved. If the B-resident measures this curvature, there arises a question: where is the curvature's center? And the answer leads beyond B into $\tilde{A}$. Even a uniform motion along a geodesic will be non-inertial with respect to that center. This implies the existence of force $\mathbf{F}_B$.

---

[1] Such distinction does not apply to 2 geodesics connecting a pair of points on a sphere, since they are the parts of one closed geodesic.

## 6. Navigations along the parallels

Once we accept $\tilde{A}$ as a legitimate conjecture, we consider another thought experiment. Let Ben travel on sphere B along a geographic parallel $L(\theta)$ of radius $r = R\sin\theta$, with latitude $\chi = (\pi/2) - \theta$ (Fig.4a). We will observe this motion from the viewpoint of Alan – resident of "Flatland" A, now intersecting B along $L$. Thus, circle $L$ is a common element of Euclidean plane A and spherical surface B. Therefore physics of uniform motion along $L$ observed by Alan must be also true for Ben, all the more so that now they will be put under identical conditions: their spaceships are moving together along $L$. But they have different legal status: Alan is a citizen of A, while Ben – citizen of B. Since the physics on $L$ does not depend on researchers' status, both travelers must experience the same centrifugal force equivalent to gravity pushing them away from rotational axis $OO'$ and balanced by centripetal force

$$F_A = \mu v^2 / r = \mu \Omega^2 r = \mu_0 \gamma(\Omega R \sin\theta) \Omega^2 R \sin\theta \tag{6}$$

Unlike the gravity associated with rotation, the centripetal force must have some physical source. According to jurisdictions of their worlds, each astronaut can directly measure only the force within his respective space. This does not affect Alan since $\mathbf{F}_A$ lies entirely within A. But Ben can only observe the *component* $F_\theta = F_A \cos\theta$ of $\mathbf{F}_A$ along the corresponding meridian. Assuming the same $\Omega$ for any $\theta$, Ben measures (Fig. 4b)

$$F_\theta = \frac{1}{2} \mu_0 \gamma(\Omega R \sin\theta) \Omega^2 R \sin 2\theta \;\to\; \mu_0 R \Omega^2 \begin{cases} \theta, & \theta \to 0 \quad (7a) \\ \gamma(\Omega R)\left(\theta - \dfrac{\pi}{2}\right), & \theta \to \pi/2 \quad (7b) \end{cases}$$

It is close to Alan's result near the Pole, where both worlds almost coincide. But as the experimental trials shift toward the equator, we have $\mathbf{F}_A \rightleftarrows \mathbf{F}_B$, and the component of $\mathbf{F}_A$ accessible to Ben, after passing through a maximum at a certain $\theta$, shrinks down to zero. In particular, Ben stops feeling any gravitational pressure (his 2D body is now orthogonal to it). This, however, does not make his motion inertial. In reality, the whole vector $\mathbf{F}_A$ being now equal to $\mathbf{F}_B$ and thus orthogonal to B, is just hidden from Ben. Yet its existence is manifest from indirect evidence. Alan's motion along the equatorial circle still requires the centripetal force $\mathbf{F}_A$. It must be provided by some physical source, e.g., by a gravitating mass at the center of $L$ in plane A. Since no such thing exists for Ben, the only way for him to explain his ship staying together with Alan's is to admit that B itself produces force $\mathbf{F}_B = \mathbf{F}_A$, even though he cannot feel it. But he can visualize it as a normal force on a pebble sliding along the surface of the balloon mentioned in the previous section.

One could object that conditions are not identical for the two partners since their 2D ships do not coincide and are mutually orthogonal on the equator. This is true, but it does not affect the strict correlation of their motions: they have common part – their intersection, belonging to both worlds.

We can also consider a slightly different version of this experiment, replacing the extended participants with two identical point masses, like quarks or leptons behaving as structureless particles [19]. Let two such particles $P_1$ and $P_2$ move along $L$ with the same speed $v$. Since $L$ is a common element of A and B, nothing can prevent us from considering one particle, say $P_1$, as belonging to A, and the other, $P_2$, as belonging to B. Let $L$ be a geodesic (e.g., the equator in Fig. 4). Then $P_1$, from the viewpoint of A, must be under force (5) to remain on $L$. The $P_2$, from the viewpoint of B, moves along $L$ without any observable force on it. But since both particles are identical and in identical motion, their being or not being under a force cannot depend on which particle belongs to which space. One could "exchange their citizenship", and then it would be $P_2$ under force (5). This again shows that the closed space exerts a force on any moving object. Generally, there is a force $\mathbf{F}_A$ with magnitude (6) on any B-particle moving in a circle. Its component (7) directly observable by Ben, must have some physical source. The hidden "radial" component $\mathbf{F}_R \equiv \mathbf{F}_B$ is given by

$$F_B \equiv F_A^R = F_R \sin\theta = \mu_0 \gamma(\Omega R \sin\theta) \Omega^2 R \sin^2\theta, \tag{8}$$

and is produced by B-space itself. For motion along geodesic ($\theta = \pi/2$), it becomes the only remaining component and is numerically equal to (5). We can rewrite (5) in the vector form:

$$\mathbf{F}_B = \hat{\mathbf{R}} \mu v^2 / R = \hat{\mathbf{R}} \mu \Omega^2 R, \tag{9}$$

where $\hat{\mathbf{R}}$ is a center-directed radial unit vector, which can be visualized in $\tilde{A}$. Apart from acting only on moving objects, force $\mathbf{F}_B$ is non-local – it is spread uniformly over the space. Being orthogonal to all dimensions of B, it cannot be observed in a local experiment with a single particle. But it becomes manifest, e.g., in a system of particles.

### 7. The Ehrenfest paradox in B

Here we generalize the previous case onto a continuous set of points $B'$, all moving along different parallels of B with a common angular velocity $\Omega$. Set $B'$ forms an extended RF rotating with respect to B and congruent with it. We will use it for a more detailed description of the Ehrenfest paradox from Sec. 2.3. Consider a circle $L(\theta)$ in Fig. 4. The radius of the circle in A is $r_A = R\sin\theta$, but in B we have two radii since any circle in B has two centers (in our case – O and O'). The "inner" (smaller) radius is

$$r_B = R\theta, \tag{10a}$$

and the "outer" (larger) one is

$$\bar{r}_B = R(\pi - \theta). \tag{10b}$$

Here it is sufficient to use one of the two, and we will use the inner one.

Just as the surface of a flat rotating disk in its own RF is non-Euclidean [5, 10-14], the rotating frame $B'$, albeit congruent with B, is described by non-spherical geometry. While the radii $r_B$, $\bar{r}_B$ remain unaffected by rotation, the *proper* length of $L(\theta)$ changes according to

$$2\pi r \to 2\pi r \cdot \gamma(\Omega r) \tag{11}$$

The change of geometry is connected with the time lag [10-14]

$$\Delta \tau' = -2 \frac{\gamma(\Omega r)}{c^2} \Omega A \equiv -2 \frac{\gamma(\Omega r)}{c^2} \Omega \pi r^2 \tag{12}$$

This is related to the corresponding inner area $B \equiv B(\theta)$ within $L(\theta)$ (we write it in italics in order to distinguish from space B). From $dB = 2\pi r R d\theta = 2\pi r^2 \sin\theta \, d\theta$ we find

$$B = \int_0^\theta dB = 2\pi R^2 (1 - \cos\theta) = B_M \left(1 - \sqrt{1 - \frac{A}{A_M}}\right), \tag{13a}$$

where $A_M \equiv \pi R^2$ and $B_M \equiv 2\pi R^2$. The "outer" (larger) area of the same circle $L(\theta)$ is given by

$$\bar{B} = \int_\theta^\pi dB = 2\pi R^2 (1 + \cos\theta) = B_M \left(1 + \sqrt{1 - \frac{A}{A_M}}\right) \tag{13b}$$

Both areas become equal at $\theta = \pi/2$ when $L$ is a geodesic. The status "inner"/"outer" changes to its opposite when $L$ crosses the equator to the southern hemisphere ($\theta > \pi/2$). Generally, as mentioned above, we use the inner radius $r_B$ and accordingly the inner area (13a). Then there follows

$$A = B \left(1 - \frac{1}{2} \frac{B}{B_M}\right), \tag{14}$$

and

$$\Delta \tau'(B) = -2 \frac{\gamma(\Omega r)}{c^2} \Omega B \left(1 - \frac{1}{2} \frac{B}{B_M}\right) \tag{15}$$

Expressing also the Lorentz factor in terms of $B$,

$$\gamma(\Omega r) = \left(1 - \frac{B\Omega^2}{\pi c^2} \left(1 - \frac{1}{2} \frac{B}{B_M}\right)\right)^{-1/2}, \tag{16}$$

we obtain

$$\Delta \tau' = -\frac{2}{c^2} \frac{\Omega B \left(1 - \frac{1}{2}\frac{B}{B_M}\right)}{\sqrt{1 - \frac{\Omega^2}{\pi c^2} B \left(1 - \frac{1}{2}\frac{B}{B_M}\right)}} \xrightarrow{B \to B_M} -\frac{2\pi}{c^2} \frac{\Omega R^2}{\sqrt{1 - \frac{\Omega^2 R^2}{c^2}}} \qquad (17)$$

If we extend B′ to the whole hemisphere, with Ben riding on the rim, then according to (17), the time lag does not disappear. Its expression just simplifies to the right-hand side of (17). Ben's time remains multi-valued, even though the rim is now geodesic. So a moving frame B′ is not equivalent to B in any global experiment. The Ehrenfest paradox in B, like the twin paradox, has the same resolution as in a flat space.

Note the three important points here. First, the considered collective motion of set B′ relative to B *requires correlated local forces to keep each particle on its specific circular path*. Second, the considered correlated motion along the geographic parallels is the only possible type of motion leaving B′ undistorted. And third, the conclusions do not require any references to Ã - e.g., the time lag (17) is expressed exclusively in terms of relevant characteristics of B and B′.

## 8. Navigations along the meridians

Another case of collective motion in B is a set B′ of non-interacting objects moving uniformly along meridians with one common speed $v$. The only force $\mathbf{F}_B$ on each object is not locally observable. But ironically, precisely such apparently "force-free" motion generally distorts the whole system B′. This becomes obvious when we notice that meridians converge on the poles.

Consider first 2 spaceships $B_1$ and $B_2$ on *the same* meridian at a distance $s_{12}$ from one another. Motion with common speed with engines off leaves $s_{12} = const$, so they remain at rest relative to each other.

But the situation is different when the ships are rearranged to move abreast along 2 different meridians. Start with 2 ships $B_1$ and $B_2$ at rest at the equatorial points of meridians 1 and 2 (Fig. 5). They form stationary system B. Next, introduce another pair of ships, $B'_1$ and $B'_2$, launched from the same intersections simultaneously, with equal North-bound velocities (system B′). With engines off right after the start, they will move uniformly toward the "North Pole" O′. Were the space flat, this motion would be truly inertial, and the pair $(B'_1, B'_2)$ could as well be considered as resting while the pair $(B_1, B_2)$ – as moving back from it; the initial distance $s'_{1,2}$ between $B'_1$ and $B'_2$ would remain constant. But on the sphere, distance $s'_{1,2}$ will shrink, and the ships will eventually collide at O'. Such an outcome is impossible for the objects in *inertial* motion along the parallel geodesics. This effect cannot be attributed to the tidal forces existing in a non-uniform gravitational field [4, 5, 9], since the distribution of matter and thereby the local geometry here is assumed uniform.

Let us extend system B′ to a set of equidistant non-interacting spaceships $B'_1, B'_2, ..., B'_n, ...$, all crossing the equator simultaneously at time $t = 0$ with the same speed *in one direction* along their respective meridians. There are no local forces within B. Under such conditions, the truly inertial motion of the set $B'_1, B'_2, ..., B'_n, ...$ would preserve its initial arrangement. But consulting with Fig. 5, we see that the distance $s'(t)$ between any two neighboring ships is a function of time

$$s'(t) = R\cos\chi\,\Delta\varphi = R\,\Delta\varphi\cos\Omega t, \qquad (18)$$

where $\chi(t) = (\pi/2) - \theta(t) = \Omega t$ is their changing latitude, $\Delta\varphi$ is their constant relative azimuth, and $\Omega$ is their common angular speed along the respective meridians. At $t \ll t_\Omega \equiv \pi/2\Omega$ the whole system of ships is close to inertial. But for an arbitrary $t$, the ships are moving with non-zero acceleration with respect to each other [4]:

$$a'(t) \equiv \frac{d^2 s'(t)}{dt^2} = -R\Omega^2\,\Delta\varphi\cos\Omega t = -\Omega^2 s'(t) \qquad (19)$$

This looks like acceleration of simple harmonic oscillator, but with no springs involved! Force $\mathbf{F}_B$ here acts as effective springs.

At $t = t_\Omega$ all the ships will collide at the pole. In contrast, the stationary set of ships $(B_1, B_2, ..., B_n)$ (system B) will remain unchanged forever.

Thus, system B′ is insensitive to its extension along the meridian, but is sensitive to extension along the parallel. Space B is isotropic, but its moving clone B′ is anisotropic (this is yet another criterion distinguishing B from B′). The only origin of such anisotropy under given conditions is *the motion itself*, which singles out its direction. And since the observed properties of B and B′ are different, the systems are not equivalent, even though their relative motion is uniform and all parts of B′ are moving along initially parallel geodesics.

There are 3 conclusions from this. First, system B′ is, in contrast to B, generally not inertial. E.g., when North-bound, it expands from the South Pole and shrinks to the North Pole, with the polar points singled out by the initial conditions. It approximates inertial system only within a sufficiently small region of space-time, and within the equatorial strip. Second, there is a state of absolute rest in B-space. And third, space B exerts a force $\mathbf{F}_B$ on any *moving* object even if the motion is along a geodesic. This force is manifest in global experiments of the type described above.

We come to a criterion defining the state of rest and an inertial RF in B: first we assume that B itself is stationary. Then a truly inertial system of *arbitrary* size in B can be only the one at rest relative to B. This requirement is met if any configuration of non-interacting particles not subjected to any external forces remains unchanged. Unlike the time lag measurement with huge waiting time in a small spatial region, now we need a sufficiently large spatial region. But in both cases, the results are obtained without any reference to anything beyond B. The angular distances $\Delta\varphi$ and $\Delta\theta$ between any two points in B are determined by the respective arcs $s_\varphi$ and $s_\theta$ and radius $R$, which itself is defined internally as $S/2\pi$ from the equatorial length $S$. And the conclusions are consistent with those obtained in the previous sections.

### 10. Geometrodynamics in B

### 10.1 *Motion in B under elastic forces*

According to Eq. (18), even non-interacting objects $(B'_j, B'_{j+1})$ in common uniform motion along their respective meridians behave as an oscillator on a spring. In order to see the dynamic properties of B, we will now connect all of them with actual springs and consider the behavior of resulting system $(B'_1, B'_2, ..., B'_n, ...)$. At the equator the springs are relaxed and have length $s_0$. As the ships advance northward, each spring gets compressed by

$$\Delta s(t) = s_0 - s'(t) = R\Delta\varphi(1 - \cos\Omega t) = 2R\Delta\varphi \sin^2 \frac{\Omega t}{2} \qquad (20)$$

If spring's reaction is described by the Hook law with the spring constant $k$, it will acquire potential energy

$$\Delta U(t) = 2k(R\Delta\varphi)^2 \sin^4 \frac{\Omega t}{2} \qquad (21)$$

The latter can only come at the cost of the kinetic energy of the ships. The circular system of compressed springs along the line $L$ in the northern hemisphere exerts on the ships south-pointing forces along the meridians. If the springs are sufficiently strong, they will eventually stop the ships at some parallel $L_\chi$ and reverse the initial motion back to the equator. In the absence of dissipation, the system will cross the equator and stop at the parallel $L_{-\chi}$ in the southern hemisphere, and the collective harmonic oscillations of the whole system will ensue.

In A, such situation could occur only for the ships converging to the center of a circle along the radial lines. "Marching" of A-ships with common speed along the straight parallel lines will not affect their connecting springs and thus will be equivalent to their state of rest.

In contrast, the considered marching of ships in B affects the connecting springs and through them – their longitudinal motion. This is another signature of the closed space, seen within B without invoking $\tilde{A}$. But the dynamics of B itself provokes us to do it when dealing with at least two moving objects.

A similar experiment in B with $D=3$ would involve a system of ships $B'_{ij}$ forming an initially flat grid moving perpendicular to itself. If the ships are autonomous, they will all collide simultaneously at a distant point – a Pole for the given arrangement. If they are connected by the initially relaxed springs, they will return, pass to the opposite semi-space, and return back for a new cycle. The grid will start oscillating along the direction perpendicular to its plane – without any observable forces acting on it. Each ship will oscillate along its respective geodesic, and all geodesics are parallel in the initial arrangement. Even with the zero net force on the grid, one would not call it an inertial system when in motion. Motion and rest would be physically different states.

### 10.2 *Motion under gravity*

Consider first two stationary charged particles $Q_1$ and $Q_2$ on some geodesic. Whatever is the analog of Coulomb's law in a closed 2*D* space, it cannot be written as one term. We will have two terms associated with two distances between the particles:

$$F_{1,2}(s) = K_B Q_1 Q_2 \left[ f(s_1) - f(s_2) \right] \qquad (22a)$$

Here $K_B$ is the analog of the Coulomb's constant in B, $s_1$ and $s_2$ are, respectively, the smallest and largest of the two separations between the particles on geodesic, and $f(s)$ is a decreasing function of *s*. We define the direction of the electric field of a charge as pointing down the shortest arc from it times the sign of the charge. When the charges are at the antipodal points, Nature itself does not know what must be the direction of force on either charge, and Eq. (22a) cleverly gives then the zero net force, as stated already in Sec. 1. It is easy to see that Gauss' law in B is not universal in B. It is a good approximation within a small region $\delta \ll R$ around a charge, but fails at distances comparable with *R*.

The similar results are true for gravitational interactions as well. For two stationary point masses $\mu_1$ and $\mu_2$ we can write:

$$F_{1,2}(s) = G_B \mu_1 \mu_2 \left[ f(s_1) - f(s_2) \right] \qquad (22b)$$

Here $G_B$ is the analog of the gravitational constant in B, and $s_1, s_2, f(s)$ have the same meaning as in (22a). The only difference is that the direction of force on $\mu_2$ *always* points to $\mu_1$ along the shortest arc between them, and vice versa, due to universal attraction.

Consider now the mass "smeared out" uniformly along the whole geodesic. Even if the linear mass density is arbitrarily high, the net force on each length element $\Delta s$ is zero within B. But the curvature of B brings the whole geodesic under tension. Like a stretched elastic string, it tends to contract. The same happens with uniform 2*D* massive sphere. The longitudinal or surface tension due to gravity produces radial forces $\mathbf{F}_B^G$ compressing the B-world. Force $\mathbf{F}_B^G$ is different from $\mathbf{F}_B$ discussed in the previous sections, but it also emerges as another attribute of B. It is force $\mathbf{F}_B^G$ that would slow down the cosmological expansion (and accelerate the contraction) of a closed Universe. Models of open and oscillating universe [4, 5, 16] ) were widely discussed before the discovery of cosmological inflation and the dark energy [20]. Even though the corresponding solutions for uniform world have no center when describing an open space, they single out the effective center for B. But again, such center lies beyond B.

### 10.3 *The center of mass*

Define the center of mass (CM) for a system of *N* point masses $\mu_j$ as usual

$$\mathbf{S}_C \equiv \frac{\sum_{j=1}^{N} \mu_j \mathbf{s}_j}{\sum_{j=1}^{N} \mu_j} \qquad (23)$$

Within B, each $\mathbf{s}_j$ is a directed arc along the respective geodesic from a chosen origin to $\mu_j$. It is easy to see that CM of a system in B is not a single-valued characteristic.

1) Start with two equal masses $\mu_1 = \mu_2$ at antipodal points. Taking these points as the North and South Pole as in Fig. 4, we see that definition (23) leads to indeterminacy – any equatorial point can be the CM, depending on choice of meridian connecting the masses.

A more general case $\mu_1 \neq \mu_2$ with $\mu_1$ at O admits, in view of (10), CM anywhere on the parallel $L(\theta)$ with

$$\theta = \pi \frac{\mu_1}{\mu_1 + \mu_2} \tag{24}$$

2) Consider the reciprocal case of uniformly distributed mass along $L$ itself. Since any circle in B has two centers, the massive $L$ has two CM – at O and O', respectively. Indeed, expression (23) modified for a continuous distribution with linear mass density $\lambda = const$, gives

$$\mathbf{S}_C = \frac{\lambda \int_L \mathbf{s}(\theta,\varphi) dL}{\lambda \int_{S(\theta)} dL} = \frac{\int_L \mathbf{s}(\theta,\varphi) dL}{2\pi R \sin\theta}, \tag{25}$$

where $\mathbf{s}(\theta,\varphi)$ is the position arc-vector of a point $(\theta,\varphi)$ on $L$. With the origin taken at one of the Poles in Fig. 4, and with the ordinary rules for addition of the opposite vectors, (25) gives $\mathbf{S}_C = 0$ for either choice, which means that CM of $L$ has two different (antipodal) positions (this is why we call such case *reciprocal* to the case with two antipodal masses).

Physically, it is natural to choose the center of the *smaller* area enclosed by $L$ as $L$'s CM. But mathematically, either choice is legal. And in case $\theta = \pi/2$ both polar points can claim the status of CM with equal right. This suggests a formal algorithm quantifying the "weights" of the two competing candidates for CM of $L$ with an arbitrary $\theta$. Represent each "weight" as a fictitious mass at the respective center, and let each fictitious mass be inversely proportional to the corresponding radius of $L$ in B. The reason for such definition is that in the limit of $L$ shrinking to the point at one of the Poles, its CM will indisputably be at that Pole (its weight will go to infinity), whereas the "weight" of the opposite CM will shrink down to a minimum. Denoting the fictitious masses at O and O' as M1 and M2, respectively, consulting with Fig. 4, and using (10a, b) we can write

$$M1 = \frac{M_0}{\pi - \theta}; \quad M2 = \frac{M_0}{\theta} \tag{26a}$$

or

$$\frac{M1}{M2} = \frac{\theta}{\pi - \theta} \tag{26b}$$

Let us now replace (M1, M2) with two real masses $\mu_1$ and $\mu_2$ such that $\mu_1 = M1$ and $\mu_2 = M2$, and solve (26b) for $\theta$. This will take us back to (24), that is, the CMs of the pair $(\mu_1, \mu_2)$ will form a continuous set along $L$. Thus, the uniform massive $L$ has two antipodal CM with "weights" $M_1$, $M_2$; and the pair of real antipodal masses $(\mu_1, \mu_2)$ determined by

(26) has CM anywhere on *L*. These two systems with reciprocal distributions can be called *dual*. A discrete system of equal equidistant point masses $\mathcal{M}_j$ on *L* will also be dual to $(\mu_1, \mu_2)$. The simplest case of such systems is $(\mu_1, \mu_2)$ at O, O', and $\mathcal{M}_1 = \mathcal{M}_2$ at any two opposite points on $L(\theta)$ with $\theta$ determined by (23). The same result is seen directly from Fig. 4 if we consider, e.g., two opposite point masses $\mathcal{M}_1 = \mathcal{M}_2$ on *L*. This system has two CMs – one in the middle O' of the smaller arc of the meridian passing through $\mathcal{M}_1$ and $\mathcal{M}_2$, with the weight M2, and the other in the middle O of the larger arc with the weight M1. The weights of M1 and M2 are given by (26).

If the B-residents look for a *single-valued* (or *effective*) CM in all considered cases, it will unavoidably lead *beyond* B, specifically – to the *interior* of spherical surface B in Fig. 4. This conclusion holds for any extended mass distribution in B.

### 10.4 *Linear momentum*

If the B-twins still insist on the absence of any forces on objects uniformly moving along geodesics, they will face another paradox in the following thought experiment. Two identical spaceships $B_1$ and $B_2$ are launched with the same speed *v* along the same meridian but in the opposite directions from the South Pole of B (Fig. 6a). The magnitude of the linear momentum of either ship is

$$P_1 = P_2 = \mu v \equiv \mu_0 \gamma(v) v \ , \qquad (27)$$

but the initial net momentum of the whole system is zero:

$$\mathbf{P} = \mathbf{P}_1 + \mathbf{P}_2 = 0 \qquad (28)$$

And it will be zero when both ships reach the North Pole. But when they cross the equator, the $\mathbf{P}_1$ and $\mathbf{P}_2$ are parallel, and the net momentum is

$$\mathbf{P} = \mathbf{P}_1 + \mathbf{P}_2 = 2\mu v \hat{\mathbf{v}}, \qquad (29)$$

where $\hat{\mathbf{v}}$ is the unit velocity-vector. Strictly speaking, this result falls short of being definitive since the vectors are now separated. Comparing two *separated* B-vectors requires bringing them close together by parallel transport, keeping components of the transported vector along the respective B-geodesics unchanged [4, 5, 21, 22]. The result depends on chosen path. Transporting one vector to the other along the initial meridian would give the initial condition (28) – the vectors would be recorded as anti-parallel even though they both point northward when separated as in Fig. 6a. However, the parallel transport along the equator confirms their parallelism and thereby Eq. (29). Such a discrepancy of mutual orientations in B is a spatial analog of the time lag in a rotating frame.

Already at this point, the conservation of momentum is shattered. The zero net momentum of the system when at the Poles turns out to be a double-valued characteristic, either (28) or (29), on the equator. The B-residents could argue that conservation law in this case dictates the appropriate choice of the transport only along the meridian, and discards the equatorial transport as illegal. But this is not a sound argument. First, both paths are mathematically legal, and there is nothing in the initial conditions that would make eligible only one of them. Second, right after the ships start from the South Pole, there appears a competing CM at the North Pole. As the ships progress northward, the priority (weight) gradually shifts from the first to the second CM as described in the previous section. When the ships meet at the North Pole, their CM is unambiguously there and nowhere else.

Thus, the north-bound motion of the system involves a concurrent north-bound motion of its *effective* CM. Any motion of CM means non-zero net momentum. Decreeing the initial momentum (28) to remain zero throughout the journey would cancel the journey. Absence of any forces in the process would lead to contradiction between conservation law and the required motion of CM. In other words, there is a clash between the claimed inertiality of uniform motion along a geodesic in B and one of the most fundamental laws of physics. The only way to avoid the conflict is to admit the reality of force (9) produced by the curvature of B-space. Such force determines evolution of **P** between (28) and (29). Applying (9) to either ship, we obtain

$$\frac{d\mathbf{P}_j}{dt} = \hat{\mathbf{R}} \mu v^2 / R = \hat{\mathbf{R}} \mu \Omega^2 R, \quad j = 1, 2 \qquad (30)$$

Force $\mathbf{F}_B$ increases with velocity $v$ and mass $\mu$ of a moving object. As $v \to c$, the $\mu \to \infty$, and there arises an interesting question whether there exists an upper limit for $\mathbf{F}_B$ in such cases. This question is in some respects similar to a problem discussed in [23]. In the current framework, we can give a tentative answer on the qualitative level that $\mathbf{F}_B$ is a flexible characteristic depending on geometry. The latter, including radius $R$, is determined by the average mass density $\rho_B$ in B. As $\mu \to \infty$, the necessary energy pumped into $\mu$ must come from the environment. This leads to decrease of the uniformly distributed $\rho_B$ and thereby increase of $R$ and the corresponding decrease of $\Omega$. Thus, B adjusts to changing conditions so as to keep any moving object, no matter how massive, within B. Accordingly, $\mathbf{F}_B$ is always sufficient to prevent the object's disappearance by slipping out into $\tilde{A}$. This is just another manifestation of conservation of matter and energy. In the extreme limit the mass density can be approximated as $\rho(\mathbf{s}, t) = \rho_B + \mu \delta(\mathbf{s} - \mathbf{v}t)$, with mass $\mu$ converted into a black hole moving nearly at the speed $c$. If it sucks in so much energy that $\rho_B$ drops below the critical value, then B gets open, $B \to A$, and the question of how to hold the rushing super-massive object within just becomes immaterial. The detailed description of such process requires the exact solution of Einstein's equations for the corresponding mass distribution.

It was argued in [14, 15] that geometry has dynamic origin. Here the dynamics seems to originate from geometry. One does not contradict the other since geometry itself is, according to GR, determined by the energy-momentum tensor.

The situation is more subtle for moving objects with angular momentum **S** as shown in Fig. 6b. Quantum-mechanically, only one component of **S** is determined in any state. Even if we consider **S** classically as a directed vector, actually it is an anti-symmetric 2-d rank tensor in 3*D* space [5, 9]. Physically, traveling of a spinning object along a curved path may affect direction of **S** (the Thomas-Wigner rotation) without any torques on it. It is a purely relativistic local effect accumulating with traveled distance [24-26]. The net effect for a closed path (the Thomas precession) must occur in B as well, even if the path is a geodesic. A detailed discussion of the corresponding problem may be the subject of a separate work.

## 11. Conclusions

1) All the above examples show the need to revise the concept of inertiality. The absence of local forces is generally not sufficient for a system moving along geodesic to be inertial. Concept of inertiality in B works within a sufficiently small region, but fails as the region increases. The only way for a system with an arbitrary size to become truly inertial in B is to remain at rest. B-space itself defines a state of absolute rest and absolute motion.

2) The absolute motion in B is manifest, e.g., through the time lag which is observed even in motion along a geodesic, and trough the force $\mathbf{F}_B$ exerted on any *moving* object. This force, being orthogonal to all dimensions of B, is undetectable in local experiments.

3) The physical phenomena in B point at the embedding Euclidean space as a useful geometrical reference. The reason for this is rather simple. At the core of the concept of curvature lies deviation of a studied curve from the tangential straight line. So the straight line is *defined* and *used* as a reference background. Even though curvature can be measured independently [21, 22, 25], the measured property *is* deviation from a straight line! Incorporating a curve even with the status of geodesic into an appropriate B-space cannot eliminate its difference from a straight line.

4) Summarizing, we can formulate what can be called "Statement B":

*One of the most fundamental principles of Physics – relativity of motion – is severely restricted in B*. Any B-space has an intrinsic scaling parameter $R$ which is absent in A. Just as the scaling parameter $c$ in velocity space restricts Galileo-Newtonian kinematics to $v \ll c$, the scaling parameter $R$ in coordinate space restricts Einstein's special relativity (SR) to regions $\delta \ll R$. For regions comparable with $R$, deviations from SR accumulate and become observable. Globally, there resurrects non-equivalence of rest and uniform motion, and the medieval Jin of absolute rest is, at least partially, back out of the box.

Thus, *a compact topology determines a preferred RF*. A space-time with quasi-static B-space is locally Lorentzian and globally – Ptolemaic.

*Acknowledgements*

I am grateful to Anwar Shiekh for turning my attention to the closed spaces, for inspiring discussions and valuable comments.

**Figures**

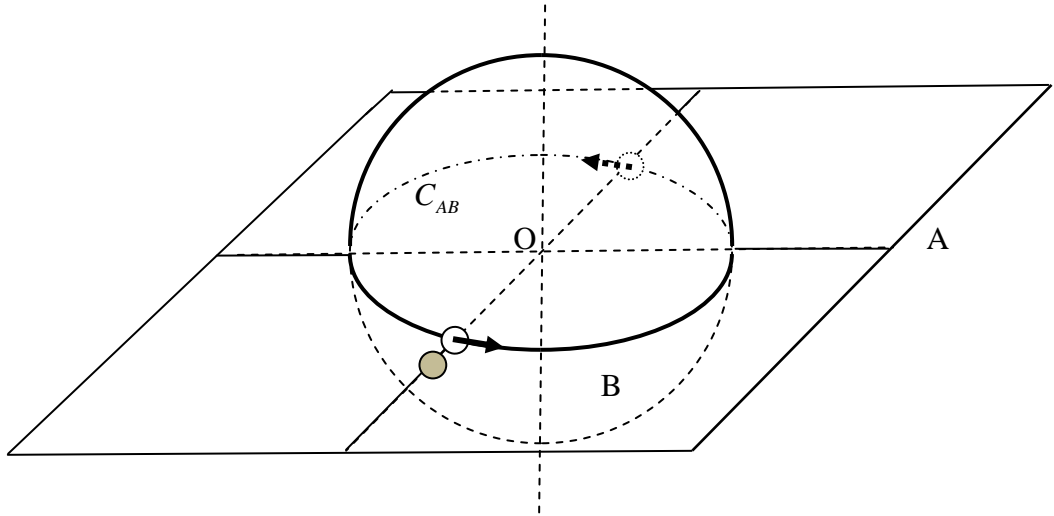

**Fig. 1**
Uniform motion along geodesic on a sphere.
Alan's spaceship (shaded circle) is stationary, while Ben's one is moving uniformly along a geodesic (equator $C_{AB}$) in a closed 2D space B. Since his ship's engines remain idle, his RF seems to be equivalent to that of Alan's, which would make the twin paradox irresolvable. But Ben's path is not geodesic in the equatorial plane A. In it, there has to be a centripetal force on his ship. This fact cannot be canceled by eliminating A. We are forced to conclude that the curvature of a closed space produces a dynamic effect equivalent to physical force on a moving object.

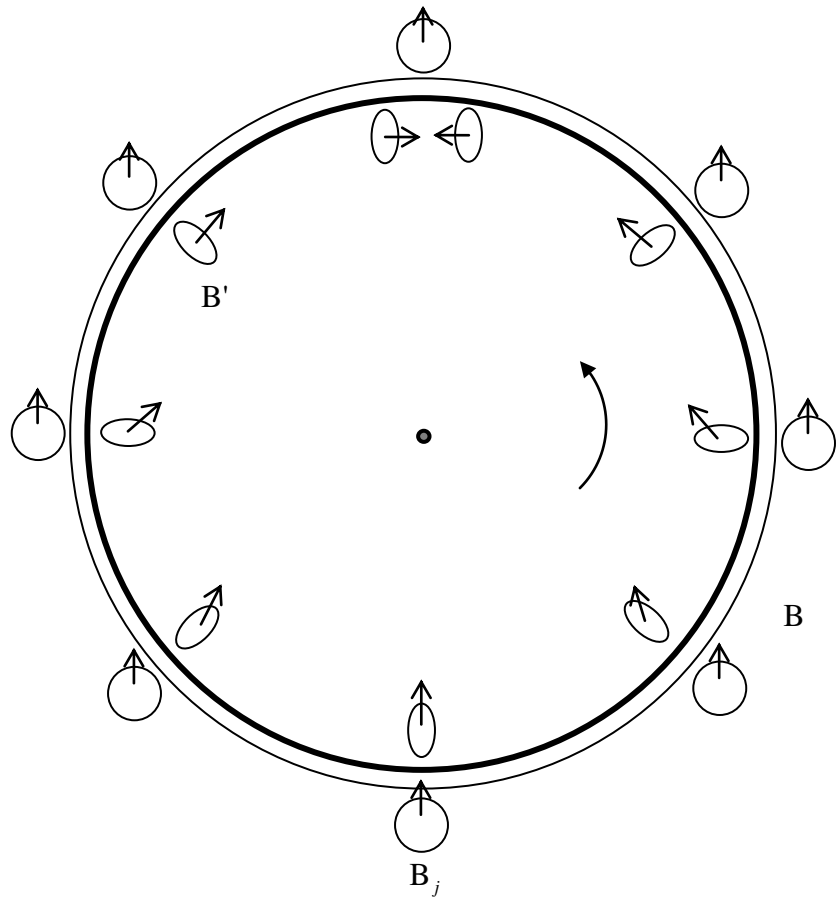

**Fig. 2**
Two sets of clocks along a circular geodesic of a closed space.
The arrows indicate the clock hands. The set B is stationary and synchronized, set B′ is moving, forming a rotational system. Its motion is manifest in the Lorentz-contraction of each individual clock as observed from B. If we synchronize set B′, going step by step around the circle, we realize that at least two different moments of time must be assigned to any event (and by iteration, we get an infinite set of equidistant moments associated with each event). This phenomenon is known as the time lag in a rotating frame. If the hand's lap around the clock dial corresponds to 1h, the time lag shown here is (1/2) h.

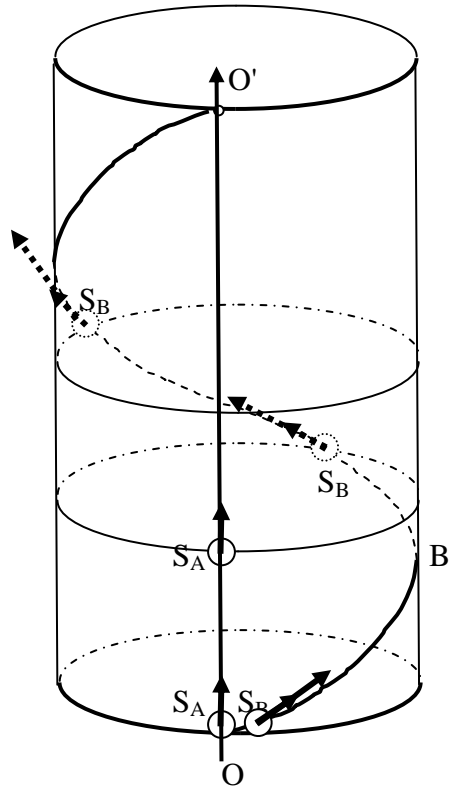

**Fig. 3**
Uniform motions along two different geodesics in a semi-closed space.
Spaceship $S_A$ is moving along a generatrix OO' of the cylindrical surface;
Spaceship $S_B$ is moving along the helix OBO'. Both ships start simultaneously
at O and meet simultaneously at O'.

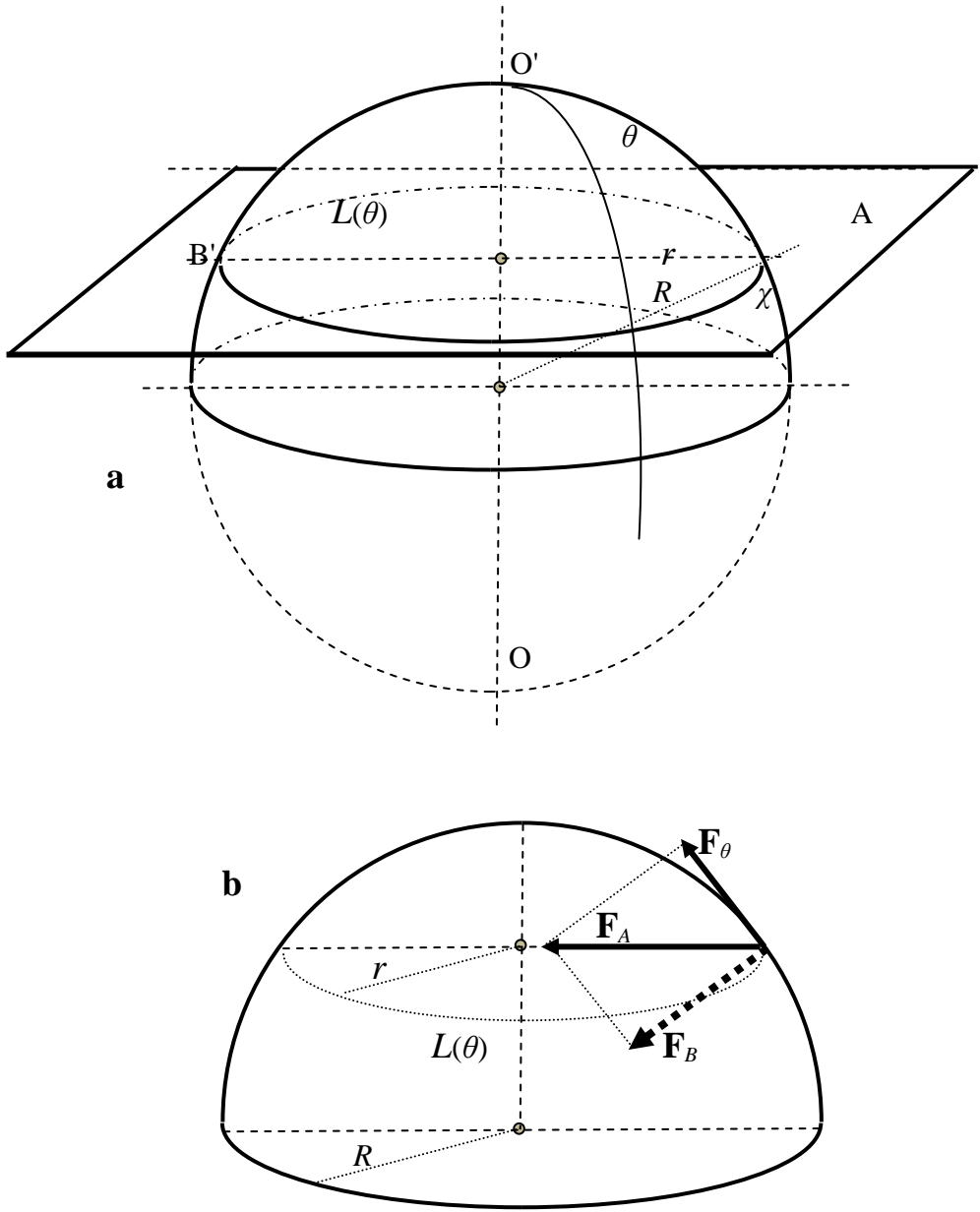

**Fig. 4**
(a) Traveling along geographic parallel $L(\theta)$ - the intersection of B and plane A.
Here $\theta$ is the polar angle, $\chi$ is the latitude, $R$ is the radius of B, and $r$ – radius of $L(\theta)$.
(b) The diagram of forces on an object moving along $L(\theta)$ in B.
 The vector sum of observable $\mathbf{F}_\theta$ and locally unobservable $\mathbf{F}_B$ gives force $\mathbf{F}_A$ in A .
 We have $(\mathbf{F}_B, \mathbf{F}_\theta) \to (0, \mathbf{F}_A)$ at $\theta \to 0$ and $(\mathbf{F}_B, \mathbf{F}_\theta) \to (\mathbf{F}_A, 0)$ at $\theta \to \pi/2$

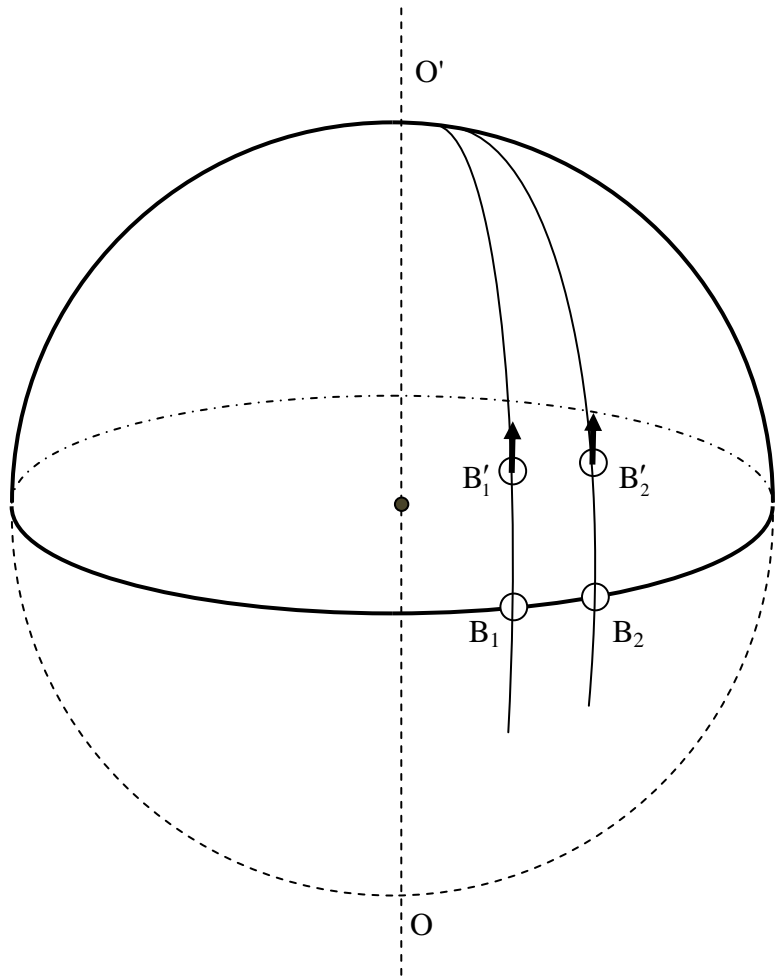

**Fig. 5**
Collective uniform motion of set B' along meridians
Only two ships $B'_1$ and $B'_2$ of the moving set are shown.
The pair $B_1$, $B_2$ represents the stationary set B.

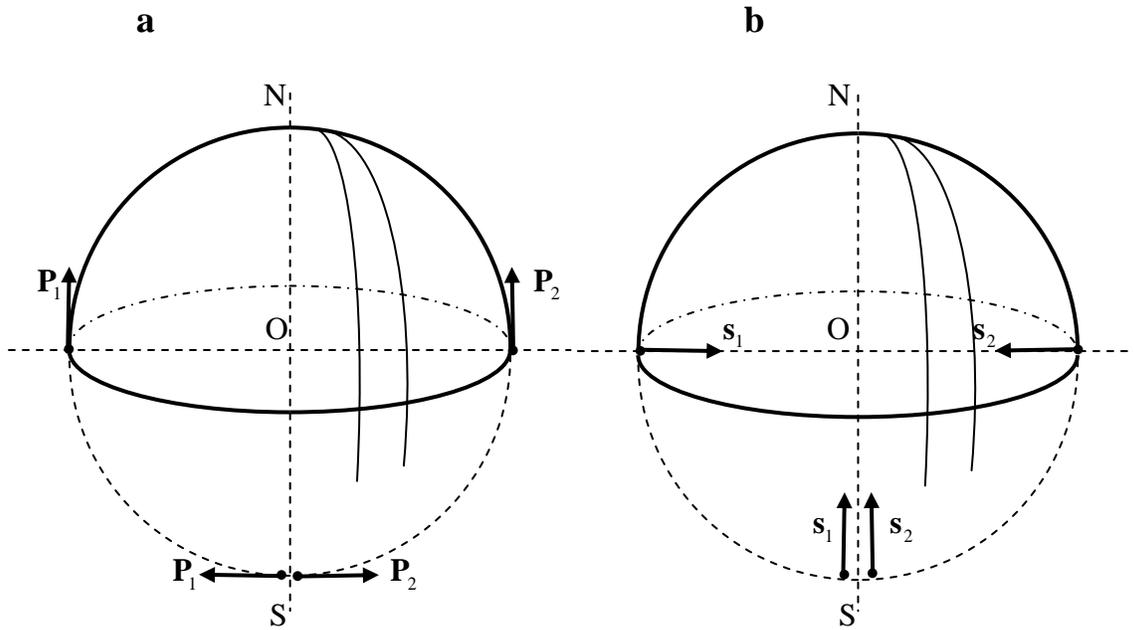

**Fig. 6**
Linear **(a)** and angular **(b)** momenta of two objects moving along the same meridian.
(a) The two objects starting at S with opposite velocities and zero net momentum **P**.
   But $\mathbf{P} \neq 0$ when they are antipodal (are passing the equator) if $\mathbf{P}_1$ and $\mathbf{P}_2$ are compared
   by the parallel transport along the equatorial line.
(b) The objects' angular momenta $\mathbf{s}_1$ and $\mathbf{s}_2$ are initially parallel when they pass each other
   at S, but may be opposite when they are antipodal.